\documentclass[%
 reprint,
 amsmath,amssymb,
 aps,
]{revtex4-2}
\usepackage{xcolor}
\usepackage{graphicx}
\usepackage{dcolumn}
\usepackage{bm}

\newcommand{\dStot}{\Delta S_{tot}}
\newcommand{\dSsys}{\Delta S_{sys}}
\newcommand{\dSint}{\Delta S_{int}}
\newcommand{\effCarnot}{\eta_{\mathrm{C}}}

\begin{document}

\preprint{APS/123-QED}

\title{Inferring entropy production from short experiments}

\author{Sreekanth K Manikandan\textsuperscript{1}}
 \author{Deepak Gupta\textsuperscript{2}}
 \author{Supriya Krishnamurthy\textsuperscript{1}}%
\affiliation{\textsuperscript{1}Department of Physics, Stockholm University, SE-10691 Stockholm, Sweden}%
\affiliation{\textsuperscript{2}Dipartimento di Fisica `G. Galilei', INFN, Universit\`a di Padova, Via Marzolo 8, 35131 Padova, Italy}

\date{\today}

\begin{abstract}
We provide a strategy for the exact inference of the average as well as the fluctuations of the entropy production in non-equilibrium systems in the steady state, from the measurements of arbitrary current fluctuations. Our results are built upon the finite-time generalization of the thermodynamic uncertainty relation, and require only very short time series data from experiments. We illustrate our results with exact and numerical solutions for two colloidal heat engines.
\end{abstract}

\maketitle

A fundamental property of non-equilibrium systems is the existence of currents which are fueled by a non vanishing average rate of  total entropy production $\sigma:= \langle \dStot\rangle/\tau$, where $\tau$ is the time  interval over which we observe  the system. 
An estimate of $\sigma$ quantifies how much heat is dissipated to the environment on average or how much free energy is lost per unit time. More information is available from  fluctuations of $\dStot$. These are governed by the fluctuation theorems \cite{Evans-Cohen,Evans-Searles,Searles-Evans,Searles2001,Gallavotti-Cohen,Gallavotti1995,Kurchan,Lebowitz-Spohn,JarzynskiPRL,crooks,Seifert:2012stf,Seifert:2005epa} and can be used for the estimation of free energy differences and \cite{hummer2001free,liphardt2002equilibrium} studying the binding energies in  single-molecule \cite{Mossa_2009, ritort2006single} or multi-molecular experiments \cite{camunas:ebe}. An accurate  quantification of the statistics of $\dStot$ could also help improve our understanding of the non-equilibrium physics of active matter systems \cite{Ramaswamy:MSA}, biological systems  \cite{Cates_2012,seara2018entropy} and nanoscale devices \cite{Pietzonka:2018utp,Manikandan:EF,Verley:2014uce,Verley:2014ute} such as colloidal heat engines \cite{Martinez:2017che,Martinez:2016bce}. 

The main challenge in the thermodynamic characterization of microscopic systems continues to be however, the lack of a  general scheme for the  measurement and characterization of $\dStot$. For systems such as colloidal particles, for which the full dynamical equations are known, {\it{stochastic thermodynamics}} provides a framework to quantify $\dStot$ from individual trajectories \cite{sekimoto1997kinetic,sekimoto1998langevin,Seifert:2012stf}. For more complex systems where not all relevant mesostates are accessible, these direct  strategies fail \cite{Seifert:inf,martinez2019inferring}. The only options are either to perform local calorimetric measurements to directly measure the heat emitted to the bath \cite{basta2018sensitive} or to come up with a new scheme for inferring $\sigma$ indirectly.

Recently, for non-equilibrium systems in a steady state, such a  scheme for identifying $\sigma$ has been proposed \cite{Seifert:inf,Gingrich:qua}  using the  thermodynamic uncertainty relation \cite{PhysRevLett.114.158101,barato:ubo}. Using this scheme, a lower bound $\sigma_L$ for $\sigma$ can be  obtained from the measurement of {\em any} fluctuating current $J$, in terms of its mean $\left\langle J \right\rangle$ and variance $\text{Var}(J)$ as,
\begin{align}
\label{eq:unc}
   \sigma \geq\sigma_L \equiv \frac{2 k_B \left\langle J \right\rangle ^2 }{\tau \text{Var}(J)}.
\end{align}
Here, $k_B$ is the Boltzmann constant. 
Eq.\ \eqref{eq:unc} holds for arbitrary $\tau$ for non-equilibrium systems in a steady state \cite{Seifert:ftg}, and the proof follows from a $\sigma$-dependent parabolic bound on the large deviation function \cite{Touchette:2009lda} of $J$ \cite{Ginrich:dba,gingrich:pft}. 

This inference scheme for $\sigma$ has been shown \cite{Gingrich:qua} to perform better than more direct methods that use spatial or temporal averages. 
However,  since the uncertainty relation is an inequality, Eq.\ \eqref{eq:unc}  still only gives a bound for $\sigma$ even when $ J=\dStot$.
How tight this bound is depends in general on model details and the $J$ chosen.
As a result, there has been much interest recently on how
to choose $J$ such that the bound value is the tightest \cite{Gingrich:qua, Pigolotti:hac}. For $\tau \rightarrow \infty$, it is known that the current 
$J$ that gives the best bound is  $J=\dStot$ \cite{Ginrich:dba}.

Eq.\ \eqref{eq:unc} could be used to predict $\sigma$ exactly if the equality was to hold.
One case when this is known to happen is the
equilibrium limit \cite{PhysRevLett.114.158101,Ginrich:dba,Manikandan:2018erf} with $J=\dStot$.  This means, that for systems working in the close-to-equilibrium/linear response regimes, there is a possibility to estimate $\sigma$ arbitrarily close to the exact value by using Eq.\ \eqref{eq:unc}. 
The equality in Eq.\ \eqref{eq:unc} is also met for arbitrary non-equilibrium conditions if along with $J=\dStot$, certain conditions are met by the steady state current and probability distributions \cite{hasegawa:pre,Physical}. However, 
there is no general scheme available so far for inferring sigma  exactly 
under arbitrary non-equilibrium conditions. In addition no scheme exists, to our knowledge, for inferring fluctuations in $\dStot$.

We address precisely these issues in this Letter. Our
first central contribution is to provide a new strategy which, in principle, can estimate $\sigma$ exactly at arbitrary non-equilibrium conditions, by using Eq.\ \eqref{eq:unc} in the $\tau\rightarrow 0$ limit. In this limit, for the current $J=\dStot$, it can be shown that 
the equality condition holds, just as for the equilibrium limit.
Using this feature, we show that we can infer $\sigma$ arbitrarily close to the exact value, by evaluating $\sigma_L$ for a variety of $J$ calculated over very short time durations, and then choosing the largest  value of $\sigma_L$ that results. 
A particularly appealing point from an experimental perspective is that, since we need to only compute $\sigma_L$ of Eq.\ \eqref{eq:unc} over very short trajectories, a single long time-series could suffice to obtain both $\langle J\rangle$ and Var$(J)$. Notice that the value of $\sigma$ so inferred is the steady-state value and hence valid at all times.

Our second contribution is to demonstrate that, by combining the value of $\sigma$ inferred from the previous step and the structure of the large deviation function of arbitrary currents \cite{Ginrich:dba,Pietzonka:ubo,gingrich:pft}, we can also infer the distribution of $\dStot$, and as a result all the cumulants, arbitrarily close to their exact values. We illustrate all our findings using exact and numerical solutions for two models of colloidal engines, namely the Brownian gyrator \cite{Filliger:2007bgm,Manikandan:EF} as well as the isothermal work-to-work converter engine \cite{Gupta:2017sei}.

We begin by considering the uncertainty relation for  $J=\dStot$, which reads (setting $k_B =1$),
\begin{align}
   \frac{\text{Var}(\dStot)}{\left\langle \dStot \right\rangle}\geq 2.
\label{eq:unc1}
\end{align}
To motivate that this inequality saturates at $\tau \rightarrow 0$, we consider the arbitrary time, scaled cumulant generating function (SCGF)  $\phi_{\dStot}(\lambda,\tau)\equiv\frac{1}{\tau}\log\left\langle e^{-\lambda\dStot}\right\rangle_\tau$. 
For short time durations, when $\vert \dStot \vert \ll 1$, we can express $\phi_{\dStot}(\lambda,\tau)$ as a series expansion in terms of the cumulants of $\dStot$. Then, to the leading order that respects convexity, we get (see the Supplemental Material \cite{suppl} for more details and explicit expressions for the first four cumulants for an exactly solvable model),
\begin{align}
\label{eq:series1}
\phi_{\dStot}(\lambda,\tau)&\sim -\frac{\lambda\left\langle\dStot\right\rangle}{\tau}+\frac{\lambda^2\text{Var}(\dStot)}{2\;\tau}.
\end{align}
Now applying the integral fluctuation theorem \cite{Seifert:2005epa}: $\phi_{\dStot}(1,\tau)=0$, we get,
\begin{align}
\label{eq:limitds}
    \frac{\text{Var}(\dStot)}{\left\langle \dStot \right\rangle}\to 2\quad\quad \text{as}\quad \quad \tau\to 0.
\end{align}
We stress that the limit $\tau \rightarrow 0$ is crucial for Eq.\ \eqref{eq:series1} and Eq.\ \eqref{eq:limitds} to hold. For an arbitrary $\tau $ they are valid \textit{only if} the distribution of $\dStot$ is a Gaussian. A more rigorous proof is provided in \cite{hasegawa:turfc} to the effect that the equality in Eq.\ \eqref{eq:unc1} is always attained when $\dStot\rightarrow 0$. However, note for our purposes, that this  happens not only  for the equilibrium limit but also when $\tau\rightarrow 0$. A model which can be solved exactly for the LHS of  Eq.\ \eqref{eq:unc1} has also been shown \cite{Manikandan:2018erf} to display this behaviour as $\tau \rightarrow 0$.  
\begin{figure}
\centering
\includegraphics[scale=0.329]{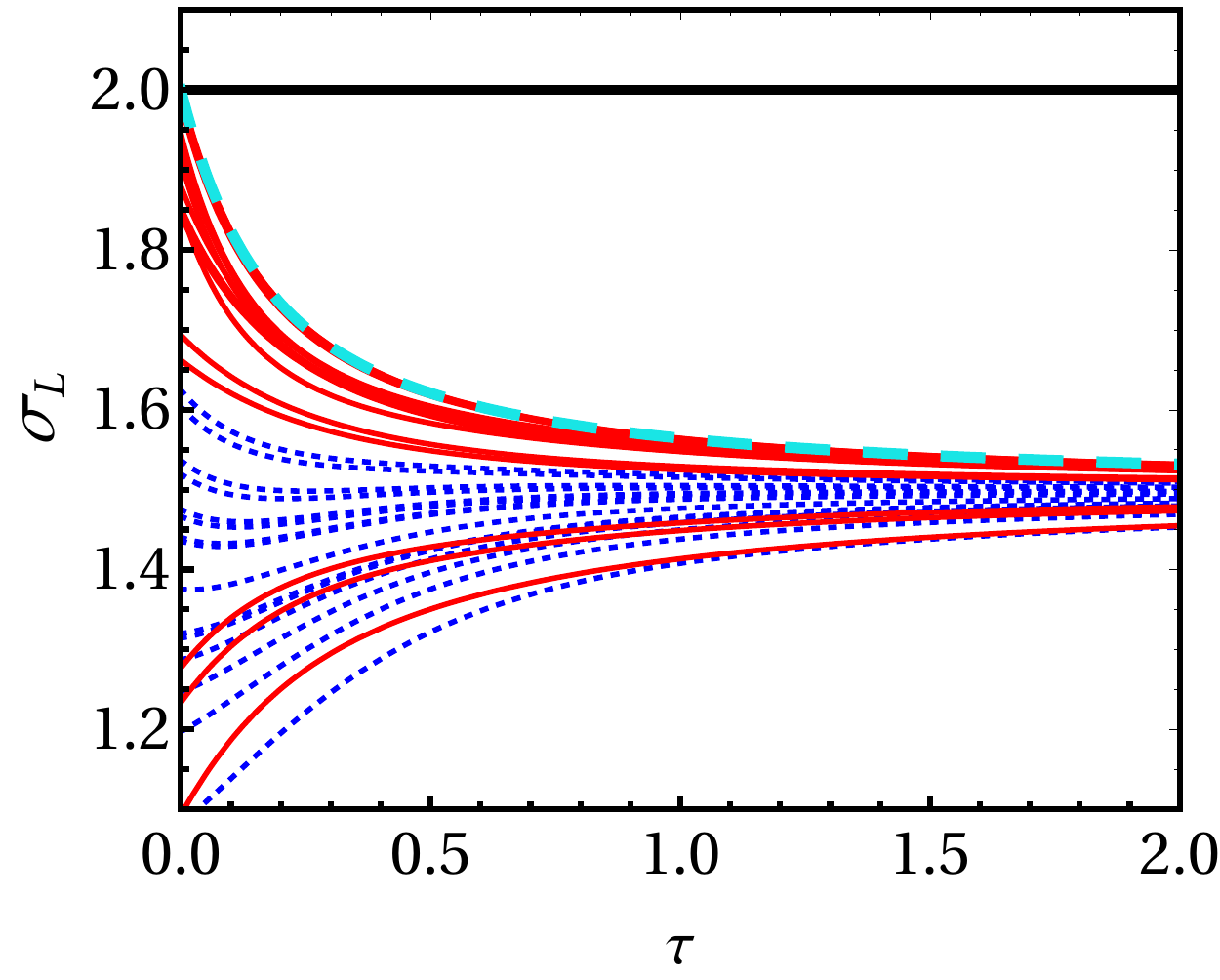}~
\includegraphics[scale=0.329]{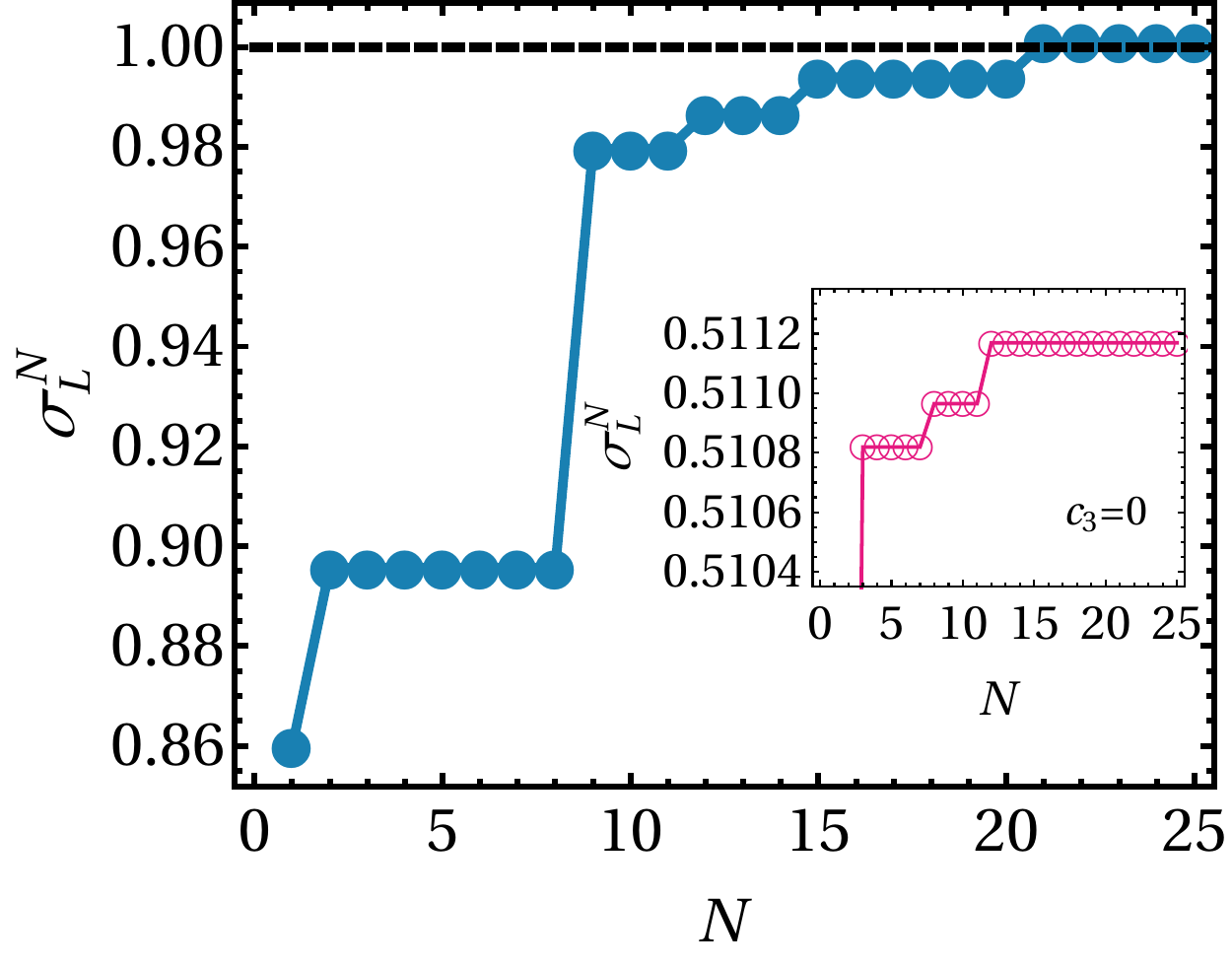}
\caption{An illustration of the exact estimation of the entropy production rate $\sigma$, using the $\tau\rightarrow 0$ limit of Eq.\ \eqref{eq:unc} for two colloidal engine models in non-equilibrium steady states. 
\textit{Left}: $\sigma$ inferred as a function of time, for the Brownian gyrator model, with analytic solutions for $\sigma_L$ in Eq.\ \eqref{eq:unc}. The black horizontal line corresponds to the actual entropy production rate. The red solid and blue dotted lines correspond to arbitrary currents with and without boundary contributions (see the main text). 
The results show that the best inference of $\sigma$ is given by $\dStot$ (green dashed line) itself, and that several of the currents having boundary contributions (red solid lines) infer $\sigma$ arbitrarily close to the actual value, in the $\tau \rightarrow 0$ limit. \textit{Right}: Inferring $\sigma$, as the maximum of the measured $\sigma_L$'s of $N$ arbitrary currents (Eq.\ \eqref{eq:fts1}), for the isothermal work-to-work converter engine, from numerical simulations. Here the black dashed line corresponds to the actual $\sigma$, obtainable from a large-time computation \cite{Gupta:2017sei}. $\sigma_L^N$ corresponds to the maximum $\sigma_L$ inferred by the $N$ currents, at $\tau \rightarrow 0$. We see that as $N$ increases, $\sigma_L^N$ saturates to the known value of $\sigma$. The inset shows that the inference procedure makes a large error if the boundary terms are not included.
\label{fig:inf2}}
\end{figure}

We now demonstrate the usefulness of Eq.\ \eqref{eq:limitds} for inferring $\sigma$ for two non-trivial models of colloidal engines, the Brownian gyrator model \cite{Filliger:2007bgm,Argun:2017erm,Manikandan:EF} and the work-to-work converter engine \cite{Gupta:2017sei,Manikandan:EF}, in both of which the working substance is a single colloidal particle. In the first case, the particle is in contact with external reservoirs at hot ($T_1$) and cold ($T_2$) temperatures and in the second case, the particle is subjected to two white-noise forces, interpreted as a load and drive force. 
In these models, there is a clear division of $\dStot$ in terms of time-extensive quantities such as heat ($Q$) and work ($W$) and time-intensive contributions such as the entropy change of the system $\dSsys$. Hence $\dStot$ can be written as,
\begin{align}
\label{eq:StotSplit}
\begin{split}
	\dStot &=\frac{\effCarnot}{T_2} Q_1 + \frac{1}{T_2} W + \dSint.
\, 
\end{split}
\end{align}
Here $\eta_C = 1-\frac{T_2}{T_1}$, is the Carnot efficiency.  

The term $\dSint = -\frac{1}{T_2} \Delta E + \dSsys$
collects the time-intensive contributions to the total entropy production that depend only on the initial and final states of the system. $\Delta E$ denotes the change in internal energy, which is,
according to the First Law, $\Delta E = W + Q_1 + Q_2$.  
We define an arbitrary current $J$ in the system as the linear combination $ J=c_1\;\frac{\eta_C}{T_2}Q_1+ c_2\;\frac{1}{T_2}W + c_3\;\dSint$, where $c_1,\;c_2$ and $c_3$ are random real numbers, taken uniformly from the interval $\left[-1,\;1\right]$. In particular, when $c_1=c_2=c_3=1$, we get $J=\dStot$ \footnote{In principle, one can also consider any arbitrary decomposition of $\dStot$ with more components. We will stick to a three dimensional basis for simplicity.}.

In Fig.\ \ref{fig:inf2} we illustrate this inference scheme for the Brownian gyrator  (Fig.\ \ref{fig:inf2}, left panel)  and the work-to-work converter engine (Fig.\ \ref{fig:inf2}, right panel).  Since both of these models
have been extensively studied in the literature, we relegate their
detailed description to the Supplemental Material \cite{suppl}.
The Brownian gyrator can be solved exactly \cite{Manikandan:EF} for the full SCGF $\Phi(\lambda_Q,\lambda_w,\lambda_S,\tau)\equiv \frac{1}{\tau}\log\left\langle e^{-\lambda_Q Q_1 - \lambda_W W + \lambda_S \dSint}\right\rangle_\tau$ at arbitrary times and hence provides us with
the means to check the inference procedure analytically (\textcolor{black}{see the Supplemental Material \cite{suppl}, which also contains the references \cite{Manikandan:2017awd,Chernyak:2006pia,Onsager:1953fai,Machlup:1953fai,Kirsten:2003fdc}} ). 
The second model of the work-to-work converter  can only be solved
for large times \cite{Gupta:2017sei}.  We hence  use it to test our inference scheme in a situation where we can only rely on numerics.

In the left panel of  Fig.\ \ref{fig:inf2}, we compute $\sigma_L$  for the Brownian gyrator, using our analytical solutions (see \cite{suppl} for more details)  for arbitrary currents
$ J$ 
at any time $\tau$. The exact value of $\sigma$ is marked by the black horizontal line. 
In the $\tau \rightarrow 0$ limit, $\dStot$ infers $\sigma$ exactly. Also, notice that there are other currents which perform almost as good as $\dStot$, and infer $\sigma$ arbitrarily close to the actual value, in the $\tau\rightarrow 0$ limit. The red solid lines correspond to a value of $\sigma_L$ computed from currents for which $c_3 \neq 0$.  The blue dotted lines correspond to $\sigma_L$ calculated from currents for which $c_3=0$ and hence which are only linear combinations of $Q_1$ and $W$, the time-extensive contributions to $\dStot$. 
The best inference strategy is therefore to measure the mean and variance of an ensemble of randomly generated currents,  at short times. 
Since, the bound in Eq.\ \eqref{eq:unc} saturates for $\tau\rightarrow 0$, we are guaranteed to obtain a value for $\sigma_L$ arbitrarily close to the actual $\sigma$ as,
\begin{align}
    \label{eq:fts1}
    \sigma=\max_{J} \big\lbrace\lim_{\tau \rightarrow 0}\sigma_L\big\rbrace.
\end{align}
Note that for large $\tau$ all currents, including $J=\dStot$ give a similar estimate, which is considerably less than the actual value (left panel, Fig.\ \ref{fig:inf2}). Hence, the small-time saturation of Eq.\ \eqref{eq:unc} as well as it's sensitivity to the chosen $J$, both work in favour of getting a better estimate for $\sigma$
than at large $\tau$.
In practice, the $\tau \rightarrow 0$ limit may be achieved in experiments by
choosing trajectory lengths corresponding to the minimal temporal resolution 
accessible to the experiment \cite{kheifets2014observation,di2014probing} such that $\dStot$ also becomes arbitrarily small in this limit.
In the right panel of Fig.\ \ref{fig:inf2}, we numerically compute $\sigma_L$ for the second model of the work-to-work converter, by computing the mean and variance of different randomly chosen $J$ for very short trajectories and 
using Eq.\ \eqref{eq:fts1}. 
Since this model can be solved in the steady state \cite{Gupta:2017sei}, the exact value of $\sigma$ is known. As can be seen, the inferred value is in very good agreement with the exact value after the inference procedure has been applied to the order of about $20$ currents. 

\begin{figure}
\centering
\includegraphics[scale=0.5]{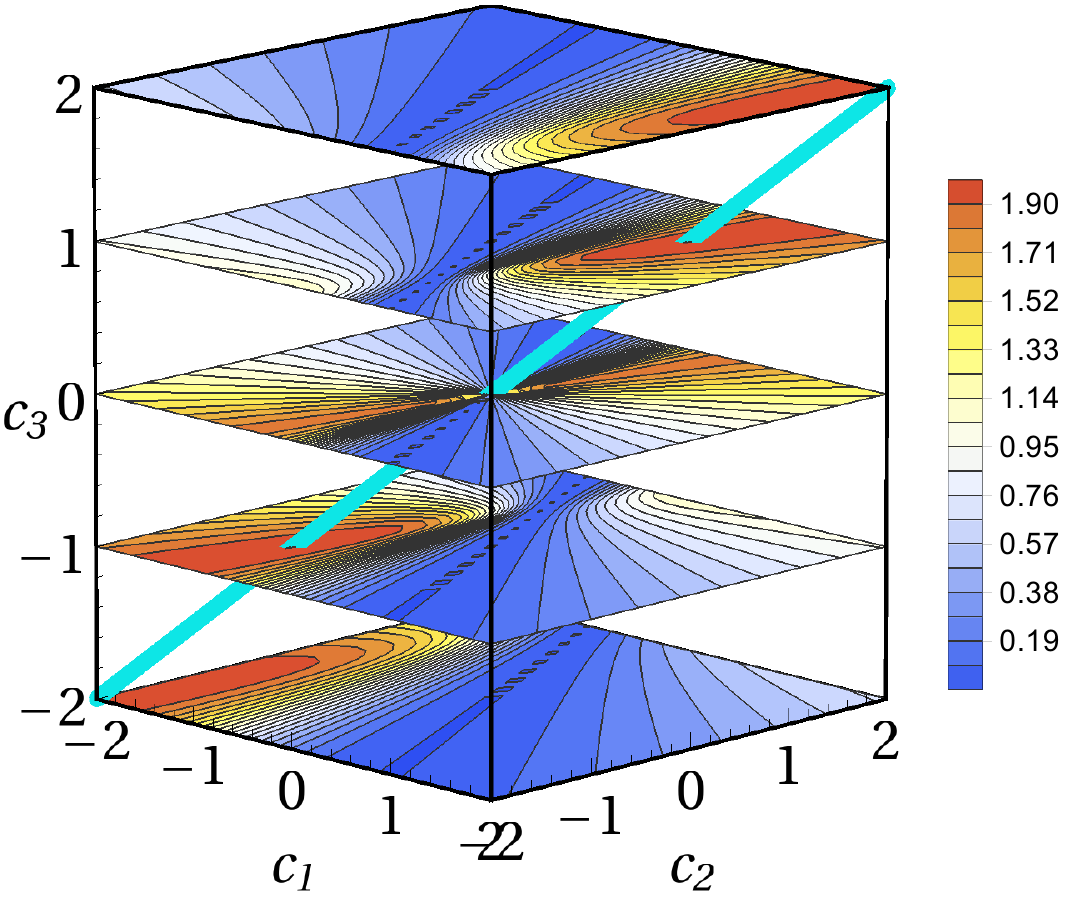}~~~~~
\caption{Entropy production rate inferred in the $\tau \rightarrow 0$ limit, for the Brownian gyrator model. Arbitrary currents are constructed as linear combinations of the basis currents $Q_1$, $W$ and $\dSint$, with coefficients $c_1$, $c_2$ and $c_3$. The green line corresponds to the $\dStot$ current direction. In the scale provided, red corresponds to a more accurate inference and blue to an inaccurate inference. It is found that, for non zero values of $c_3$, the best inference is given by the $\dStot$ current itself. However, the optimal current differs from the corresponding $\dStot$, when we set $c_3=0$ (see Supplemental Material \cite{suppl}). 
\label{fig:bestinf1}
}
\end{figure}

To further analyze the inference of $\sigma$ by the short-time inference scheme, Eq.\ \eqref{eq:fts1}, 
we have identified the optimal currents that infer $\sigma$ the best. 
In Fig.\ \ref{fig:bestinf1}, we illustrate this in the $\tau \rightarrow 0$ limit for the Brownian gyrator. When $ c_3\;\neq 0$ the current that infers best is $\dStot$ itself, as expected. However, for $c_3=0$,  
some other direction in the $(c_1$, $c_2)$ plane gives the current that infers best (Fig.\ \ref{fig:bestinf1}). 
Even when $\tau > 0$ , we find (see Supplemental Material \cite{suppl}), that the optimal current is in general different from $\dStot$ as recently suggested in \cite{Pigolotti:hac}, and becomes equal to $\dStot$  only for large $\tau$ \cite{Ginrich:dba}. 

It is important to note that, our usage of the $c_1,c_2,c_3$ -space is purely a proof of concept, to demonstrate the viability and applicability of our short-time inference scheme, as well as to disentangle the separate roles played by the time-intensive and time-extensive terms which would not have been obvious otherwise. In addition, in this space, currents have natural physical interpretations in terms of work, heat and system entropy production. For a generic non-equilibrium system however, for which neither a Langevin nor a Master equation description is known, such a decomposition of $\dStot$ (Eq.\ \eqref{eq:StotSplit}) is usually not possible. In such cases, one can generate random currents from the phase space trajectories $\textbf{x}(t)$ of the system \cite{Gingrich:inf,Gingrich:qua} using random vectors $\textbf{d}(\textbf{x})$ 
as $ J\equiv J_d=\int_{\textbf{x}(0)}^{\textbf{x}(\tau)}d\textbf{x}\;\textbf{d}(\textbf{x})\;\textbf{j}(\textbf{x})$, where  $\textbf{j}(\textbf{x})=\frac{1}{\tau}\int_0^\tau \delta[\textbf{x}-\textbf{x}(t)]\;d\textbf{x}(t)$ is an estimate for a steady state current. Such a current can then be used to compute $\sigma_L$ in Eq. (\ref{eq:unc}) and many such currents can be generated by varying $\textbf d$. Our scheme will then imply that as $\tau \rightarrow 0$, the current that will satisfy the equality in Eq. (\ref{eq:unc}) will correspond to $\dStot$. Note that $J_d$'s defined this way
contain  information of both the extensive and intensive contributions and there is no need to calculate the steady state distribution separately. In the Supplemental Material, we show the $\textbf d$ field which corresponds to $J= \dStot$ for the Brownian gyrator. For problems where an analytical treatment is ruled out, such a $\textbf d$ field would have to be obtained by some optimisation procedure \cite{Gingrich:qua}.

So far, we have shown that the finite time thermodynamic uncertainty relation can be used at very short observational times to infer $\sigma$  arbitrarily close to the actual value, arbitrarily far from equilibrium. 
It is then  natural to ask, if there exist similar inference strategies for the fluctuations  of $\dStot$ as well. Since the current that infers $\sigma$ the best at $\tau = 0$ is $J=\dStot$, a naive solution would have been to approximate the optimal current in the $\tau \rightarrow 0$ limit by $\dStot$, and compute a distribution directly from that. Although this can work in principle, there are a couple of practical disadvantages. First, with the real data, one is always limited by the minimal experimental resolution $\Delta t > 0$ and the $\tau \rightarrow 0$ limit can be accessed only within this error. Secondly, there can be several near optimal currents (as shown in Figure.\ \ref{fig:inf2}, \textit{Left}) that infer $\sigma$ arbitrarily close to the exact value, which a computational scheme may not be able to distinguish. For {\it e.g.,} it has been noticed in \cite{Gingrich:qua}, for finite $\tau$, that these near optimal currents can look very different from each other, and still predict a similar value for $\sigma$. As a result of these limitations, because of which we may have access to a value $\sigma$, but not the current $J=\dStot$, we present an alternate strategy for computing the steady state distribution of $\dStot$ using an
exact estimate of $\sigma$.

 We begin by considering the structure of the scaled cumulant generating function of an arbitrary current in the steady state,  $ \phi^\sigma_{J}(\lambda,\tau) \equiv\frac{1}{\tau} \log \left\langle e^{-\lambda \frac{\sigma\tau J }{\left\langle J \right\rangle}}\right\rangle_\tau$ at large $\tau$. Using large deviation techniques, it has been shown recently that $ \phi^\sigma_{J}$ obeys the bound  \cite{Pietzonka:ubo,Ginrich:dba},
\begin{align}
\label{eq:phibound}
     -\sigma\lambda(1- \lambda)\leq \phi_{\dStot}(\lambda,\tau)\leq  \phi^\sigma_{J}(\lambda,\tau),
\end{align}
The uncertainty relation in Eq.\ \eqref{eq:unc} can  be directly proved from this result \cite{Ginrich:dba,gingrich:pft}. 
Interestingly, Eq.\ \eqref{eq:phibound} constrains the fluctuations of $\dStot$ strongly, by providing both a lower bound and an upper bound for $\phi_{\dStot}$. In particular, one can saturate the bound $\phi_{\dStot}(\lambda,\tau)\leq \phi^\sigma_{J}(\lambda,\tau)$ 
if $J=\dStot$.
We therefore get a natural scheme for inferring $\phi_{\dStot}(\lambda,\tau)$ as,
\begin{align}
\label{eq:scgfinf}
    \phi_{\dStot}(\lambda,\tau)=\min_{ J }\left\lbrace \phi^\sigma_{J}(\lambda,\tau)\right \rbrace.
\end{align}
Consequently, if  $M_J^{(n)}$ is the $n$-th cumulant of the normalized current $\sigma \tau J/\langle J\rangle$, we also get (see \cite{suppl}),
\begin{align}
\label{eq:infvar}
    M_{\dStot}^{(n)}&=\min_J\lbrace M_J^{(n)}\rbrace
    .
    \end{align}
The cumulants thus inferred can then be used to construct the histogram of $\dStot$ straightforwardly. 
\begin{figure}
\centering
\includegraphics[scale=0.26]{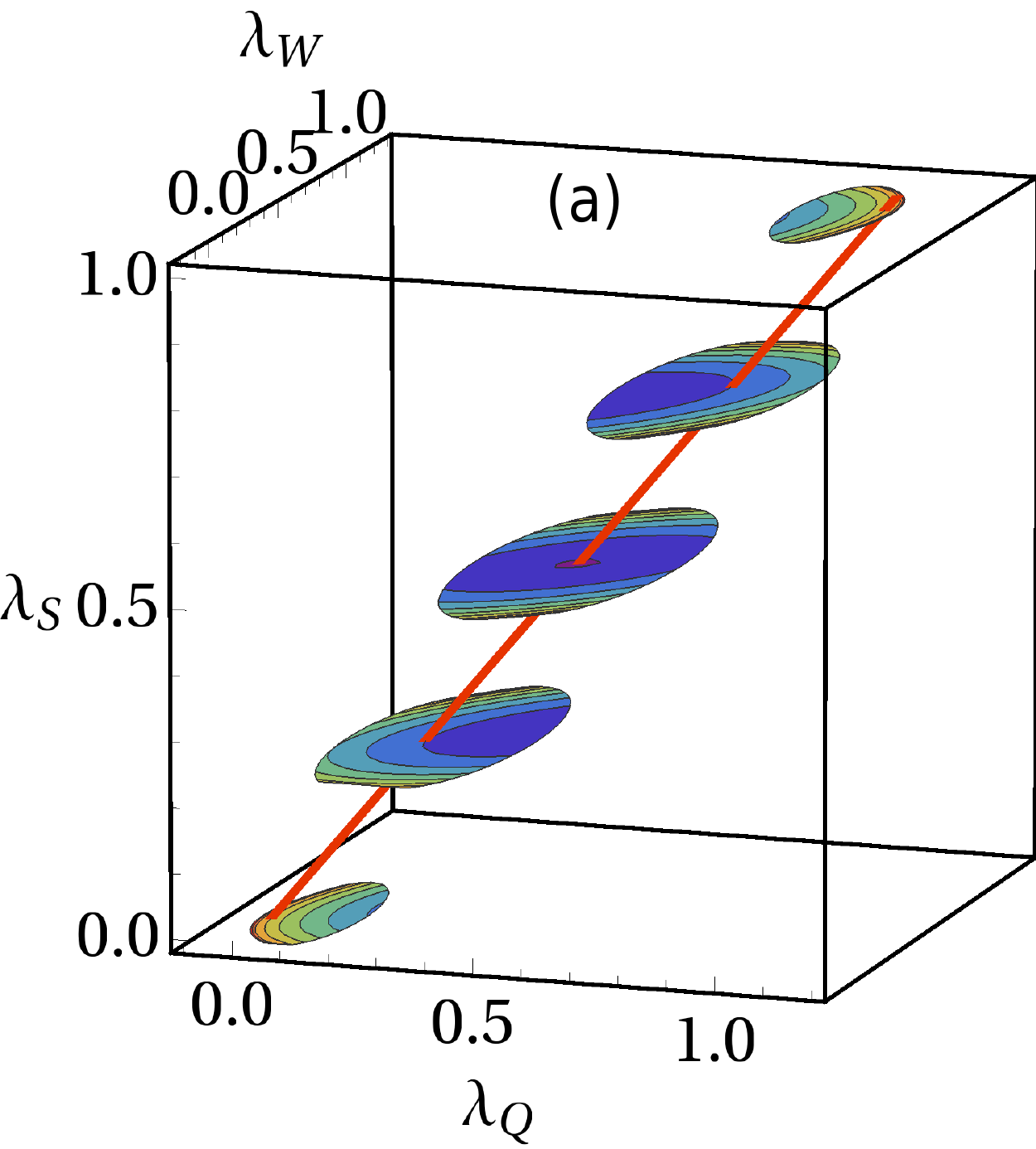}~~~
\includegraphics[scale=0.3]{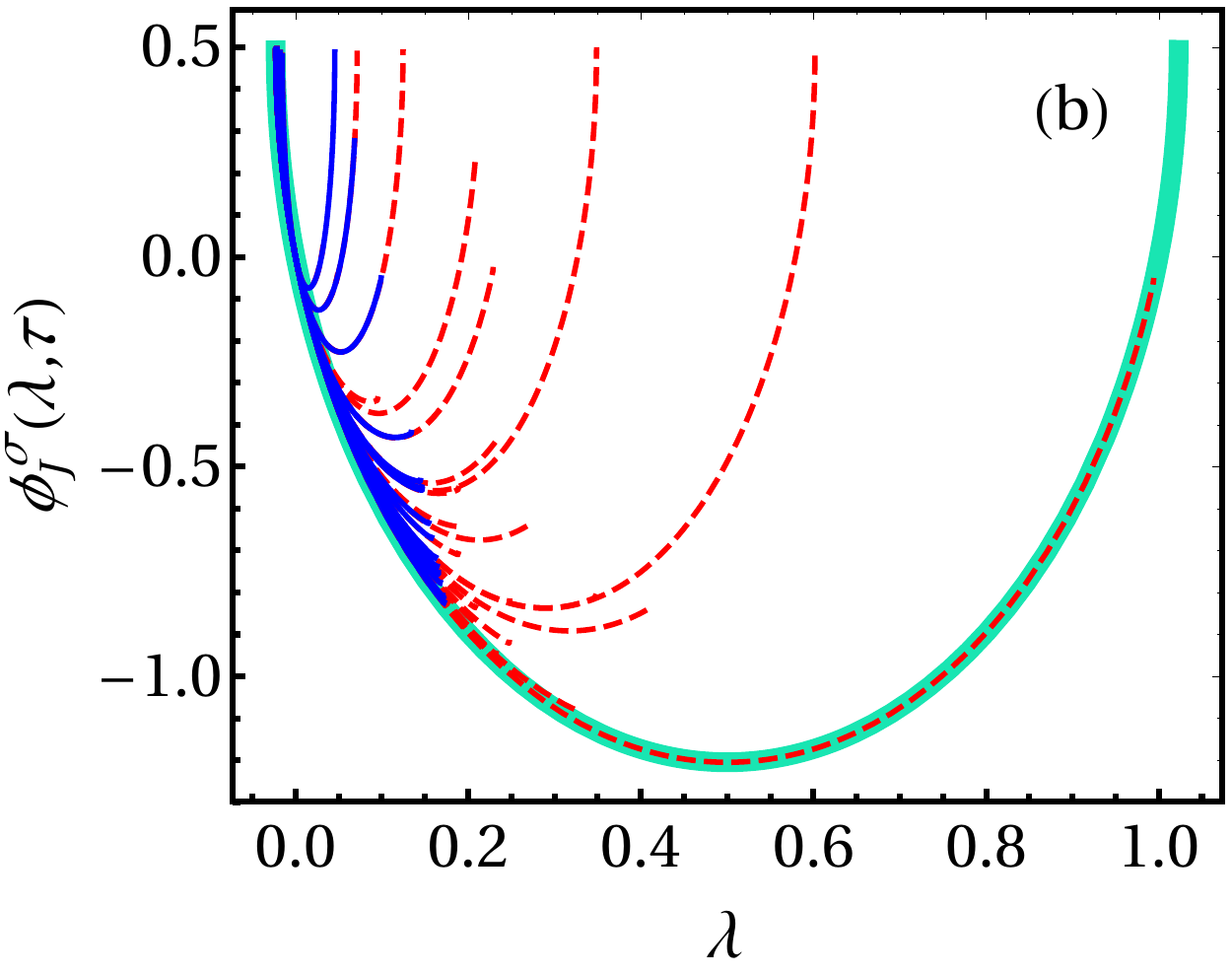}\\
\includegraphics[scale=0.3]{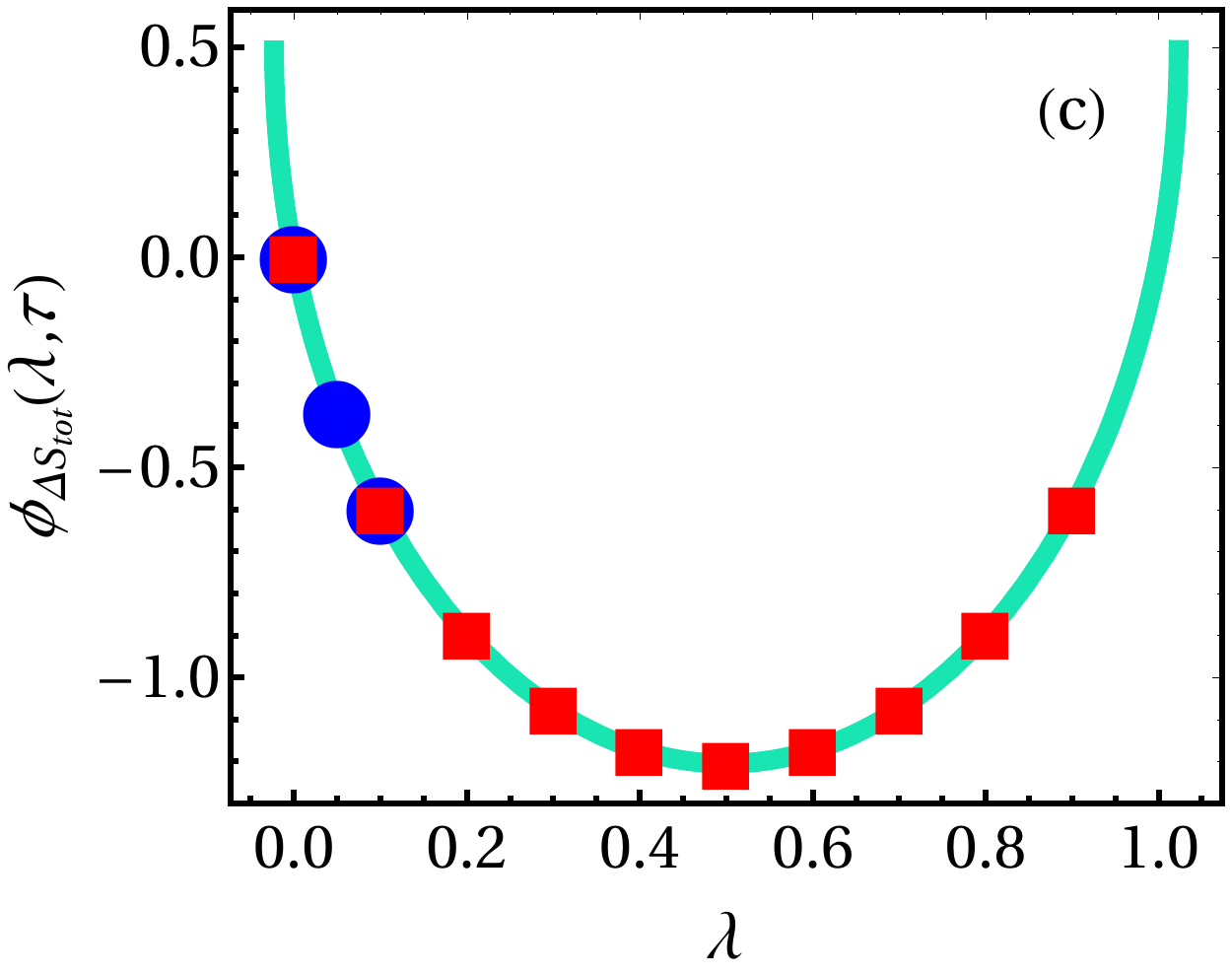}~~~
\includegraphics[scale=0.28]{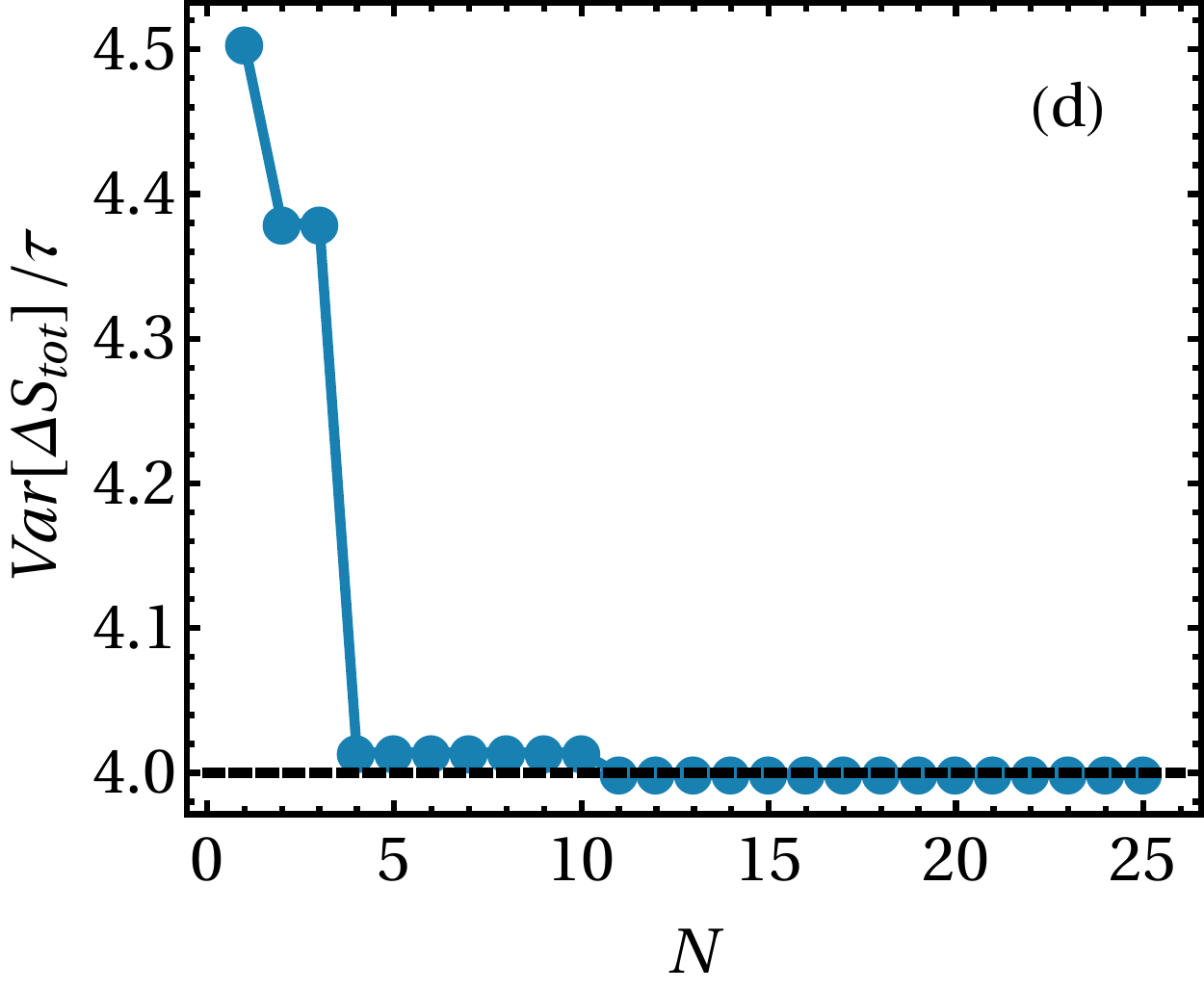}
\caption{ Results for the isothermal work-to-work converter engine. 
(a) $\Phi(\lambda_Q,\lambda_W,\lambda_S,\tau)$ obtained analytically. $\Phi$ has a limited domain of convergence (only a few $\lambda_S$-planes are shown for clarity), and also displays the fluctuation theorem symmetry around the point $(0.5,0.5,0.5)$. In (b) and (c), the thick green curve is the actual $\phi_{\dStot}$. In (b), the blue solid and red dashed curves, respectively, correspond to $\phi_J^\sigma$ with $c_3=0$ and $c_3\neq0$ case. In (c),  $\phi_{\dStot}$ is obtained using Eq.\ \eqref{eq:scgfinf}, where red squares and blue circles, respectively, correspond to $c_3 \neq0$ and $c_3=0$ at large $\tau$ \cite{suppl}. (d) Inferring Var($\dStot$), as the minimum of the measured variances of $N$ scaled, arbitrary currents, $\frac{\sigma \tau J}{\left\langle J \right\rangle}$ according to Eq.\ \eqref{eq:infvar}. Here the black dashed line corresponds to the variance of $\dStot$, in the long time limit 
\cite{Gupta:2017sei}.
\label{fig:iso}
}
\end{figure}

We illustrate Eqs.\ \eqref{eq:scgfinf} and  \eqref{eq:infvar} in Fig.\ \ref{fig:iso} for the isothermal work-to-work converter engine.
We have first obtained an analytic expression for the joint SCGF   $\Phi(\lambda_Q,\lambda_W,\lambda_S,\tau)=\frac{1}{\tau}\log\langle e^{-\frac{\lambda_Q \eta_C Q_1}{T_2} - \frac{\lambda_W W}{T_2} - \lambda_S \, \dSint}\rangle_\tau$, which is exact at large but finite times (see the Supplemental Material \cite{suppl}). The geometry of $\Phi$ was recently conjectured and discussed in some detail in \cite{Manikandan:EF}. Due to the fluctuation theorem, $\Phi$ is a reflection symmetric object around the point $(0.5,0.5,0.5)$, and typically has a limited domain of convergence (cut-offs) that depend on $\lambda_S$. We illustrate this for a fixed, large value of $\tau$ in Fig.\ \ref{fig:iso}a. The SCGF of an arbitrary current $J$ can  be  obtained from $\Phi$  by  evaluating it along a straight line passing through the origin and the point $\left( c_1,c_2,c_3 \right)$, where $c_1$, $c_2$, $c_3$ are random numbers. In particular, $\phi_{\dStot}$ is $\Phi$ evaluated along the (1,1,1) direction marked by the red solid line in Fig.\ \ref{fig:iso}a. 

In Figs.\ \ref{fig:iso}b 
and \ref{fig:iso}c, we illustrate the inference of $\phi_{\dStot}$ using Eq.\ \eqref{eq:scgfinf}. 
Since 
$ \phi^\sigma_{J}$ of currents with $c_3=0$, can have restricted domains of convergences (see $\lambda_S=0$ plane of Fig.\ \ref{fig:iso}a), they will end up inferring a limited domain of $\phi_{\dStot}$, as shown in  Fig.\ \ref{fig:iso}c with the blue circles. The red squares show the improvement in the estimate of $\phi_{\dStot}$ when $c_3\neq0$.
In Fig.\ \ref{fig:iso}d, we illustrate the inference of Var$(\dStot)$ using Eq.\ \eqref{eq:infvar}, numerically.   

In summary, we have presented a scheme to exactly infer the average entropy production rate $\sigma$ as well as the distribution of entropy production $P(\dStot)$ in non-equilibrium steady state systems. The scheme for identifying $\sigma$ is built upon the finite time thermodynamic uncertainty relation \cite{gingrich:pft,Seifert:ftg} and it's saturation in the very short time limit. 
The inference of $P(\dStot)$ is then built upon an exactly estimated value of $\sigma$ and the dissipation bounded structure of steady state current fluctuations \cite{Pietzonka:ubo,Ginrich:dba}.  

We have chosen to demonstrate our scheme using a decomposition of $\dStot$ into time-extensive and intensive contributions, but we expect that our scheme is equally applicable for currents generated any other way. Such considerations will be very relevant when testing this scheme in biological systems or active matter systems in the steady state. Generalizations to  time symmetrically driven systems \cite{hasegawa:turfc} could also be interesting.\\$ $\\
\textcolor{black}{\textit{Note added:} Recent preprints \cite{otsubo2020estimating,van2020entropy} have independently found further proofs/refinements of the short-time inference scheme, and have proposed efficient computational algorithms to implement it.  }

\begin{acknowledgments}
Deepak Gupta acknowledges the support from University of Padova  through  project  ``Excellence Project 2018'' of the Cariparo foundation. Supriya and Sreekanth thank Lennart Dabelow, Ralf Eichhorn, Shun Otsubo, Stefano Bo and Takahiro Sagawa for helpful discussions. We also thank Shun Otsubo for important comments on an earlier version of the manuscript.
\end{acknowledgments}

\end{document}